\documentclass[amssymb,aps,showpacs]{revtex4}

\begin{document}
\title{{\bf Quantum gauge boson propagators in the light front}}

\author{A. T. Suzuki$^{a}$\footnote{On sabbatical leave\\ Permanent
address: Instituto de F\'{i}sica Te\'{o}rica-UNESP, Rua Pamplona 145 -
01405-900 -
S\~{a}o Paulo, SP - Brazil.} and J.H.O. Sales$^{b}$}
\affiliation{$^{a}$Department of Physics, North Carolina State
University,\\ Raleigh, NC 27695-8202\\
$^{b}$Faculdade de Tecnologia de S\~{a}o Paulo-DEG, \\
P\c{c}a. Coronel Fernando Prestes, \\ 01124-060, S\~{a}o Paulo, SP}

\date{\today}

\begin{abstract} 
Gauge fields in the light front are traditionally addressed via the
employment of an algebraic condition $n\cdot A=0$ in the Lagrangian
density, where $A_{\mu}$ is the gauge field (Abelian or non-Abelian) and
$n^\mu$ is the external, light-like, constant vector which defines the
gauge proper. However, this condition though necessary is not
sufficient to fix the gauge completely; there still remains a residual
gauge freedom that must be addressed appropriately. To do this, we
need to define the condition $(n\cdot A)(\partial \cdot A)=0$ with
$n\cdot A=0=\partial \cdot A$. The implementation of this condition in
the theory gives rise to a gauge boson propagator (in momentum space)
leading to conspicuous non-local singularities of the type $(k\cdot
n)^{-\alpha}$ where $\alpha=1,2$. These singularities must be
conveniently treated, and by convenient we mean not only
matemathically well-defined but physically sound and meaningfull as
well.  In calculating such a propagator for one and two noncovariant
gauge bosons those singularities demand from the outset the use of a
prescription such as the Mandelstam-Leibbrandt (ML) one. We show that
the implementation of the ML prescription does not remove certain
pathologies associated with zero modes. However we present a causal,
singularity-softening prescription and show how to keep causality from
being broken without the zero mode nuisance and letting only the
propagation of physical degrees of freedom.

\end{abstract}

\maketitle


\section{The light-front ``environment''}

In 1949 Dirac \cite{dirac}, showed in his pioneering work on relativistic
dynamics that it is possible to construct dynamical forms from the description
of a initial state of a given relativistic system in any space-time surface
whose lengths between points have no causal connection. The dynamical evolution
corresponds to the system following a trajectory through the hyper-surfaces.
For example, the hyper-surface $t=0$ is our three-dimensional space. It is
invariant under translations and rotations. However, in any transformation of
inertial reference frames which involves ``boosts'', the temporal coordinate is
modified, and therefore the hyper-surface in $t=0$. Other hyper-surfaces can be
invariant under some type of ``boost''. It is the case of the hyperplane called
null plane, defined by $x^+=t+z$, which in analogy to the usual coordinate
system, is commonly referred to as the ``time'' coordinate for the front form
(light front). For example, a ``boost'' in the $z$ direction does not modify
the null plane.

A point in the usual four-dimensional space-time is defined through the set
of coordinates $\left( x^{0},x^{1},x^{2},x^{3}\right)$, where $x^{0}$ is the
time coordinate, that is, $x^{0}=t$, with the usual convention of taking the
speed of light equal to unity, $c=1$. The other coordinates are the
three-dimensional Euclidean space coordinates $x^{1}=x$, $x^{2}=y$ and $
x^{3}=z$.

The light-front coordinates are defined in terms of these by the following
relations: 
\begin{eqnarray}
x^{+}&=&x^{0}+x^{3},  \nonumber \\
x^{-}&=&x^{0}-x^{3},  \nonumber \\
\vec{x}^{\perp }&=& x^{1}\vec{i}+x^{2}\vec{j},
\end{eqnarray}
where $\vec{i}$ and $\vec{j}$ are the unit vectors in the direction of the
coordinates $x$ and $y$. The null plane is defined by $x^{+}=0$, that is,
this condition defines the hyper-surface which is tangent to the light cone,
the reason why some authors call those light-cone coordinates.

Note that for the usual four-dimensional Minkowski space-time whose metric $
g^{\mu\nu}$ is defined such that its signature is $(1,-1,-1,-1)$ we have 
\begin{eqnarray}
x^{+}&=&x^{0}+x^{3}\;\;=\;\;x_0-x_3\;\;\equiv \;\;\lambda\,x_{-},  \nonumber \\
x^{-}&=&x^{0}-x^{3}\;\;=\;\;x_0+x_3\;\;\equiv \;\;\lambda\,x_{+},  \nonumber \\
\vec{x}^{\perp }&=&
x^{1}\vec{i}+x^{2}\vec{j}\;\;=\;\;-x_{1}\vec{i}-x_2\vec{j}\;\;\equiv
\;\;-\vec{x}_{\perp},
\end{eqnarray}

We emphasize that in the extreme right-hand side of the above two
first lines we did not equate $x_0\mp x_3=x_\mp$ but there is a
normalizing factor $\lambda=2$ due to the structure of the
four-dimensional Minkowski metric translated into the light front
coordinates. In other words, $g^{+-}=2$, whereas $g_{+-}=\frac{1}{2}$. We could, of course, have symmetrized them, using, for
example:
\[x^{+}=\frac{x^{0}+x^{3}}{\sqrt 2}\;\;=\;\;\frac{x_0-x_3}{\sqrt2}\;\;\equiv
\;\;x_{-}, \]

The initial boundary conditions for the dynamics in the light front
are defined in this hyper-plane. Note that the $x^{+}$ axis is
orthogonal to the plane $x^{+}=0$. Therefore, a displacement of this
hyper-surface for $ x^{+}>0 $ is analogous to the displacement of the
plane $t=0$ for $t>0$ of the usual four-dimensional space-time. With
this analogy we identify $x^{+}$ as the {\em ``time''} coordinate for
the null plane. Of course, since there is a conspicuous discrete
symmetry between $x^+ \leftrightarrow x^-$, one could choose $x^-$ as
his ``time'' coordinate. However, once chosen, one has to stick to the
convention adopted. We shall adhere to the former one.

The canonically conjugate momenta for the coordinates $x^{+},x^{-}$
and $ x^{\perp }$ are defined respectively by:
\begin{eqnarray}
k^{+}&=&k^{0}+k^{3},  \nonumber \\
k^{-}&=&k^{0}-k^{3},  \nonumber \\
k^{\perp }&=&\left( k^{1},k^{2}\right) .  \label{1.4}
\end{eqnarray}

The scalar product in the light front coordinates becomes therefore 
\begin{equation}
a^{\mu }b_{\mu }=\frac{1}{2}\left( a^{+}b^{-}+a^{-}b^{+}\right) - \vec{a}
^{\perp }\!\!\cdot \vec{b}^{\perp },  \label{1.7}
\end{equation}
where $\vec{a}^{\perp }$ and $\vec{b}^{\perp }$ are the transverse
components of the four vectors. All four vectors, tensors and other entities
bearing space-time indices such as Dirac matrices $\gamma ^{\mu }$ can be
expressed in this new way, using components $(+,-,\perp)$.

From (\ref{1.7}) we can get the scalar product $x^{\mu}k_{\mu }$ in the
light front coordinates as: 
\begin{equation}
x^{\mu }k_{\mu }=\frac{1}{2}\left( x^{+}k^{-}+x^{-}k^{+}\right) - \vec{x}
^{\perp }\!\!\cdot \vec{k}^{\perp }.  \label{1.8}
\end{equation}

Here again, in analogy to the usual four-dimensional Minkowski space-time
where such a scalar product is 
\begin{equation}
x^{\mu} k_{\mu}=x^0 k^0-{\bf {x}\!\cdot {k}}
\end{equation}
where ${\bf x}$ is the three-dimensional vector, with the energy $k^0$
associated to the time coordinate $x^0$, we have the light-front
``energy'' $ k^-$ associated to the light-front ``time'' $x^+$. Note,
however, that there is a crucial difference between the two
formulations: while the usual four-dimensional space-time is
Minkowskian, the light-front coordinates project it onto two
sectorized two-dimensional Euclidean spaces, namely $(+,-)$, and $
(\perp,\perp)$.

In the Minkowski space described by the usual space-time coordinates we have
the relation between the rest mass and the energy for the free particle
given by $k^{\mu }k_{\mu }=m^{2}$. Using (\ref{1.7}), we have 
\[
k^{\mu }k_{\mu }=\frac{1}{2}\left( k^{+}k^{-}+k^{-}k^{+}\right) - \vec{k}
^{\perp }\cdot \vec{k}^{\perp }, 
\]
so that 
\begin{equation}
k^{-}=\frac{\vec{k}^{\perp 2 }+m^{2}}{k^{+}}.  \label{1.10}
\end{equation}

Note that the energy of a free particle is given by $k^{0}=\pm
\sqrt{m^{2}+ {\bf {k}^{2}}}$, which shows us a quadratic dependence
of $k^{0}$ with respect to ${\bf k}$. These positive/negative energy
possibilities for such a relation were the source of much difficulty
in the interpretation of the negative energy particle states in the
beginning of the quantum field theory descriprion for particles,
finally solved by the antiparticle interpretation given by Feynman. In
contrast to this, we have a linear dependence between $ (k^{+})^{-1}$
and $k^{-}$ (see Eq.(\ref{1.10})), which immediately reminds us of the
non-relativistic quantum mechanical type of relationship for one
particle state systems. However, herein one is squarely confronted
with the fact that for the light-front ``energy'' $k^-$ there is a
troublesome feature emerging: The vanishing of the $k^+$ momentum
component leads to a divergence in the energy relation. This is the
famous zero-mode problem in the light front milieu.

\section{The classical vector gauge boson propagator}

In a recent work of ours \cite{reex1}, we showed that a single Lagrange
multiplier defined by $(n\cdot A)(\partial \cdot A)$ with $n\cdot A=\partial
\cdot A=0$ at the classical level leads to a propagator in the light-front
gauge that has no residual gauge freedom left.

Thus, for the relevant gauge fixing term that enters in the Lagrangian
density which we define as 
\begin{equation}
(n\cdot A)(\partial \cdot A)=0,
\end{equation}
gives for the Abelian gauge field Lagrangian density: 
\begin{equation}
{\cal L}=-\frac{1}{4}F_{\mu \nu }F^{\mu \nu }-\frac{1}{2\alpha }\left(
2n_{\mu }A^{\mu }\partial _{\nu }A^{\nu }\right) ={\cal L}_{{\rm E}}+{\cal L}
_{GF}  \label{8}
\end{equation}
where the gauge fixing term is conveniently written so as to symmetrize the
indices $\mu $ and $\nu $, and the gauge parameter can assume complex
values. By partial integration and considering that terms which bear a total
derivative don't contribute and that surface terms vanish since $
\lim\limits_{x\rightarrow \infty }A^{\mu }(x)=0$, we have 
\begin{equation}
{\cal L}_{{\rm E}}=\frac{1}{2}A^{\mu }\left( \square g_{\mu \nu }-\partial
_{\mu }\partial _{\nu }\right) A^{\nu }  \label{9}
\end{equation}
and 
\begin{equation}
{\cal L}_{GF}=-\frac{1}{\alpha }(n\cdot A)(\partial \cdot A)=-\frac{1}{%
2\alpha }A^{\mu }\left( n_{\mu }\partial _{\nu }+n_{\nu }\partial
_{\mu}\right) A^{\nu }  \label{25}
\end{equation}
so that 
\begin{equation}
{\cal L}=\frac{1}{2}A^{\mu }\left( \square g_{\mu \nu }-\partial _{\mu
}\partial _{\nu }-\frac{1}{\alpha }(n_{\mu }\partial _{\nu }+n_{\nu
}\partial _{\mu })\right) A^{\nu }  \label{11}
\end{equation}

To find the gauge field propagator we need to find the inverse of the
operator between parenthesis in (\ref{11}). That differential operator in
momentum space is given by $O_{\mu \nu }(k)=-k^{2}g_{\mu \nu }+k_{\mu
}k_{\nu }+\frac{1}{\alpha }\left( n_{\mu }k_{\nu }+n_{\nu }k_{\mu }\right) $
, so that the propagator of the field, which we call $S^{\mu \nu }(k)$, must
satisfy the following equation $O_{\mu \nu }S^{\nu \lambda }\left( k\right)
=\delta _{\mu }^{\lambda }$, where $S^{\nu \lambda }(k)$ can now be
constructed from the most general tensor structure that can be defined,
i.e., all the possible linear combinations of the tensor elements that
composes it (the most general form includes the light-like vector $m_{\mu }$
dual to the $n_{\mu }$ \cite{progress} -- but for our present purpose it is
in fact indifferent): 
\begin{eqnarray}
G^{\mu \nu }(k) &=&g^{\mu \nu }A+k^{\mu }k^{\nu }B+k^{\mu }n^{\nu }C+n^{\mu
}k^{\nu }D+k^{\mu }m^{\nu }E+  \nonumber \\
&&+m^{\mu }k^{\nu }F+n^{\mu }n^{\nu }G+m^{\mu }m^{\nu }H+n^{\mu }m^{\nu
}I+m^{\mu }n^{\nu }J  \label{a2}
\end{eqnarray}

Then, it is a matter of straightforward algebraic manipulation to get the
relevant classical propagator in the light-front gauge, namely, 
\begin{equation}  \label{classicprop}
S^{\mu \nu }(k)=-\frac{1}{k^{2}}\left\{ g^{\mu \nu }-\frac{k^{\mu }n^{\nu
}+n^{\mu }k^{\nu }}{k^{+}}+\frac{n^{\mu }n^{\nu }}{(k^{+})^{2}}k^{2}\right\}
\,  \nonumber
\end{equation}

Note that this propagator differs from the traditional two-term
light-front propagator in that ours have a third term proportional to
$(k^+)^{-2}$, namely
\[\frac{n^{\mu }n^{\nu }}{(k^{+})^{2}}k^{2}.\]

We point out that this exact third term also appears in the canonical
quantization of gauge fields in the light-front
\cite{kogutetc}. Yet, many have been the
reasonings and arguments put forth to neglect it altogether. However,
we argue back that this term is relevant, first of all from the very
fact that such a term ensures there is no anomaly in going from the
classical to the quantum propagator and secondly because if $(k\cdot
n)^{-1}$ yields a manifestation of zero mode problem, so doubly does
the $(k\cdot n)^{-2}$ term, except that they come with opposite
signs. We emphasize that no third term such as this one can ever
appear at the classical level if one starts off from the Lagrangian
density with only one Lagrange multiplier of the form $(n\cdot A)$.

\section{The quantum vector gauge  boson propagator}

The Feynman quantum propagator for the gauge boson can be derived
integrating over all the momenta in (\ref{classicprop}). Projecting
out this propagator on to the light-front we get a gauge boson
particle propagating at equal light-front times. We are going to
restrict our calculation to the total momentum $P^{+}$ positive and
corresponding forward light-front time propagation. In this case the
propagator from $x^{+}=0$ to $x^{+}>0$ is given by:
\begin{equation}
\widetilde{S}^{(1)\mu \nu }(x_{1}^{\mu })=i\,\int \frac{d^{4}k_{1}}{\left(
2\pi \right) ^{4}}\frac{N^{\mu \nu }{\rm e}^{-ik_{1}^{\mu }x_{1\mu }}}{
k_{1}^{2}+i\varepsilon }.  \label{pc1}
\end{equation}
where 
\begin{equation}
N^{\mu \nu }=\frac{-g^{\mu \nu }k_{1}^{+2}+\left( k_{1}^{\mu }n^{\nu
}+n^{\mu }k_{1}^{\nu }\right) k_{1}^{+}-n^{\mu }n^{\nu
}k_{1}^{2}}{k_{1}^{+2}}.  \label{n}
\end{equation}

Note that because of the structure of the light-front propagator (\ref
{classicprop}) only three of the component projections are non vanishing,
namely, 
\begin{equation}
N^{\perp \perp }=-g^{\perp \perp },\qquad N^{\perp -}=\frac{
n^{-}k_{1}^{\perp }}{k_{1}^{+}}\qquad N^{--}=\frac{n^{-}n^{-}k_{1}^{\perp 2}
}{k_{1}^{+2}}  \label{cn}
\end{equation}
At equal light-front times, we have (we focus on the ``energy'' integral): 
\begin{equation}
\widetilde{S}^{(1)\mu \nu }(x^{+})=\frac{i}{2}\int \frac{dk_{1}^{-}}{\left(
2\pi \right) }\frac{N^{\mu \nu }{\rm e}^{-\frac{i}{2}k_{1}^{-}x^{+}}}{
k_{1}^{+}\left( k_{1}^{-}-\frac{(k_{1}^{\perp })^{2}}{k_{1}^{+}}+\frac{
i\varepsilon }{k_{1}^{+}}\right) }.  \label{1b2}
\end{equation}
so that, in terms of the component projections we have immediately 
\begin{eqnarray}
\widetilde{S}^{++}=\widetilde{S}^{+-}=\widetilde{S}^{+\perp } &=&0.
\label{cs} \\
\widetilde{S}^{\perp \perp }\neq \widetilde{S}^{\perp -}\neq \widetilde{S}
^{--} &\neq &0  \nonumber
\end{eqnarray}

Here the $(\perp,\perp)$ component presents no particular difficulty in
evaluation nor does it present any troublesome feature.  However, the
components $(\perp,-)$ and $(-,-)$, whose relevant computation is
technically similar to each other, come with the troublesome 
feature known as the zero mode problem in the light front. Without
loss of generality, we shall therefore restrict ourselves to the
analysis of the $(\perp,-)$ component in the following.

In terms of Fourier transform we have (we focus on the ``time'' variable): 
\begin{equation}
S^{(1)\mu \nu }(p^{-})=\int dx^{+}\widetilde{S}
^{(1)\mu \nu }(x^{+}){\rm e}^{\frac{i}{2}p^{-}x^{+}}\ ,  \label{tf}
\end{equation}
so that the component $S^{\perp -}$ is: 
\begin{eqnarray}
S^{(1)\perp -}(p^{-}) &=&i\int \frac{dk_{1}^{-}}{2(2\pi)}\frac{2(2\pi)\delta
\left( p^{-}-k_{1}^{-}\right)\,k_{1}^{\perp }\;n^{-} }{k_{1}^{+}\left( k_{1}^{-}-\frac{(k_{1}^{\perp
})^2}{k_{1}^{+}}+\frac{i\varepsilon }{k_{1}^{+}}\right) }\left[ \frac{1}{
k_{1}^{+}}\right] _{{\rm ML}}  \nonumber \\
&=&i\int dk_{1}^{-}\frac{k_{1}^{\perp }\;n^{-}\delta \left(
p^{-}-k_{1}^{-}\right) }{k_{1}^{+}\left( k_{1}^{-}-\frac{(k_{1}^{\perp })^{2}}{
k_{1}^{+}}+\frac{i\varepsilon }{k_{1}^{+}}\right) }  
 \left[ \frac{k_{1}^{-}}{k_{1}^{+}\left( k_{1}^{-}+\frac{
i\varepsilon }{k_{1}^{+}}\right) }\right] _{{\rm ML}}  \label{prlf1}
\end{eqnarray}
where 
\[
2\;\delta \left(p^{-}-k_{1}^{-}\right)
=\delta\left(\frac{p^--k_1^-}{2}\right)=\frac{1}{2\pi }\int
dx^{+}{\rm e}^{ \frac{i}{2}\left( p^{-}-k_{1}^{-}\right) x^{+}}
\]
and the index {\rm ML} stands for the Mandelstam-Leibbrandt prescription 
\cite{ml} for the treatment of the $(k^{+})^{-1}$ poles, namely, 
\begin{equation}
\left[ \frac{1}{k^{+}}\right] _{{\rm ML}}=\lim_{\varepsilon \rightarrow 0}
\left[ \frac{k^{-}}{k^{+}k^{-}+i\varepsilon }\right] _{{\rm ML}}  \label{ml}
\end{equation}

\noindent The result is: 
\begin{equation}
S^{(1)\perp -}(p^{-})=\frac{\theta(p^+)p^{\perp }n^{-}}{p^{+2}}\frac{i}{
\left( p^{-}-K_{0}^{(1)-}+i\varepsilon \right) },  \label{1b3}
\end{equation}
where we have introduced the definition 
\begin{equation}
K_{0}^{(1)-}=\frac{p^{\perp 2}}{p^{+}},  \label{1b4}
\end{equation}
as the light-front Hamiltonian of the free one-particle system. Note that
for $x^{+}<0$, the $S^{(1)}(x^{+})=0$ because $p^{+}>0$. Moreover, observe
that $S^{(1)\perp -}(p^{-})$ is in an operator form with respect to $p^{+}$
and $\vec{p}^{\perp }$. Consequently we have a clear manifestation of the
zero mode problem in the factor $(p^+)^{-2}=0$. 

We emphasize that even with the use of the Mandelstam-Leibbrandt
prescription to handle the non-local singularity, the zero mode
problem still lingers on.

\section{\noindent The two vector gauge boson propagators}

The two-boson gauge propagator can be derived from the covariant propagator
for two particles propagating at equal light-front times. Without losing
generality, we are going to restrict our calculation to the total momentum $
P^{+}$ positive and corresponding forward light-front time propagation. In
this case the propagator from $x^{+}=0$ to $x^{+}>0$ is given by: 
\begin{eqnarray}
\widetilde{S}^{(2)\mu \nu ;\alpha \beta }(x^{\prime \mu }{},x^{\mu })
&=&\int \frac{d^{4}k_{1}}{\left( 2\pi \right) ^{4}}\frac{d^{4}k_{2}}{\left(
2\pi \right) ^{4}}\frac{iN^{\mu \nu }{\rm e}^{-ik_{1}^{\mu }\left( x_{1\mu
}^{\prime }-x_{1\mu }\right) }}{k_{1}^{2}+i\varepsilon }  \nonumber \\
&&\frac{iN^{\alpha \beta }{\rm e}^{-ik_{2}^{\mu }\left( x_{2\mu }^{\prime
}-x_{2\mu }\right) }}{k_{2}^{2}+i\varepsilon }.  \nonumber
\end{eqnarray}

At equal light-front times $x_{1}^{+}=x_{2}^{+}=0$ and $x^{\prime
+}_1=x^{\prime +}_2=x^{+}$, the propagator is written as: 
\begin{equation}
\widetilde{S}^{(2)}(x^{+})=\widetilde{S}_{1}^{(1)}(x^{+})\widetilde{S}
_{2}^{(1)}(x^{+}),  \label{s2a}
\end{equation}
where the one-body propagators, $\widetilde{S}_{i}^{(1)}$, corresponding to
the light-front propagators of particles $i=1$ or $2$, are defined by Eq.(
\ref{pc1}). We have explicitly: 
\begin{eqnarray}
\widetilde{S}^{(2)\mu \nu ;\alpha \beta }(x^{+}) &=&i^2\int \frac{
dk_{1}^{-}}{2\left( 2\pi \right) }\frac{dk_{2}^{-}}{2\left( 2\pi \right) } 
\frac{N^{\mu \nu }{\rm e}^{-\frac{i}{2}k_{1}^{-}x^{+}}}{k_{1}^{+}\left( k_{1}^{-}- 
\frac{k_{1}^{\perp 2}-i\varepsilon }{k_{1}^{+}}\right) }  \nonumber \\
&\times &\frac{N^{\alpha \beta }{\rm e}^{-\frac{i}{2}k_{2}^{-}x^{+}}}{
k_{2}^{+}\left( k_{2}^{-}-\frac{k_{2}^{\perp 2}-i\varepsilon }{k_{2}^{+}}
\right) }.
\end{eqnarray}

The Fourier transform to the total light-front energy $P^{-}$ is given by 
\begin{equation}
S^{(2)\mu \nu ;\alpha \beta }(P^{-})=\int
dx^{+}\widetilde{S}^{(2)\mu \nu ;\alpha \beta }(x^{+})\,{\rm
e}^{\frac{i}{2}P^{-}x^{+}}\ .
\end{equation}

As before, we can recognize immediately that the following components
vanish (we ommit the $(2)$ index as well as the $P^{-}$ dependence for
shortness)
\begin{eqnarray}
S^{++,++} &=&S^{++,+-}=S^{++,+\perp}=S^{++,--}=S^{++,-\perp}=0\nonumber\\
S^{++,\perp \perp }&=&S^{+-,+-}=S^{+-,+\perp}=S^{+-,--}=S^{+-,-\perp}=0 \nonumber \\
S^{+-,\perp\perp}&=&S^{+\perp,+\perp}=S^{+\perp,--}=S^{+\perp,-\perp}=S^{+\perp,\perp\perp}=0
\end{eqnarray}
whereas the following ones are the ones that we need to deal with:
\begin{eqnarray} 
S^{--,--} & \neq &S^{--,-\perp }\neq S^{--,\perp \perp }\neq 0 \nonumber \\
S^{-\perp,-\perp} &\neq & S^{-\perp,\perp \perp}\neq S^{\perp \perp
,\perp \perp }\neq 0. \label{2boson}
\end{eqnarray}

Let us then evaluate the {\footnotesize$(\perp\!\!-,\perp\!\!-)$}-compomnent, 
\begin{eqnarray}
S^{(2)\perp -,\perp -}(P^{-}) &=&-\frac{1}{2\left( 2\pi \right) }\int \frac{
dk_{1}^{-}}{k_{1}^{+}\left( P^{+}-k_{1}^{+}\right) }  \label{plf2} \\
&\times &\left[ \frac{k_{1}^{-}}{k_{1}^{+}\left( k_{1}^{-}+
\frac{i\varepsilon }{k_{1}^{+}}\right) }\right] _{{\rm ML}}\frac{
n^{-}k_{1}^{\perp }}{\left( k_{1}^{-}-\frac{k_{1}^{\perp 2}}{k_{1}^{+}}+
\frac{i\varepsilon }{k_{1}^{+}}\right) }  \nonumber \\
&\times &\left[ \frac{P^{-}-k_{1}^{-}}{\left( P^{+}-k_{1}^{+}\right) \left(
P^{-}-k_{1}^{-}+\frac{i\varepsilon }{P^{+}-k_{1}^{+}}
\right) }\right] _{{\rm ML}}  \nonumber \\
&\times &\frac{n^{-}\left(P^\perp -k_{1}^{\perp }\right)}{\left( P^{-}-k_{1}^{-}-\frac{\left(
P^{\perp }-k_{1}^\perp \right) ^{2}}{P^+-k_{1}^{+}}+\frac{i\varepsilon }{P^{+}-k_{1}^{+}
}\right) }\ ,  \nonumber
\end{eqnarray}
where $P^{-,+,\perp }=k_{1}^{-,+,\perp }+k_{2}^{-,+,\perp }$.

We perform the analytical integration in the $k_{1}^{-}$ momentum by
evaluating the residues at the poles 
\begin{eqnarray*}
k_{1}^{-} &=&-\frac{i\varepsilon }{k_{1}^{+}}\nonumber \\
k_{1}^{-} &=&P^{-}+\frac{i\varepsilon }{P^{+}-k_{1}^{+}}.
\end{eqnarray*}
It implies that only $k_{1}^{+}$ in the interval $0<k_{1}^{+}<P^{+}$ gives a
nonvanishing contribution to the integration. The result is 
\begin{equation}  \label{perpm}
S^{(2)\perp -,\perp -}(P^{-})=\frac{1}{2}\frac{\theta
  (k_{1}^{+})}{k_1^{+ 2}}\frac{\theta
  (P^{+}-k_{1}^{+})}{(P^+-k_1^{+})^2}\frac{i\,(k_{1}^{\perp
  }n^{-})(P^{\perp }-k_{1}^{\perp })n^{-}}{\left(
  P^{-}-K_{0}^{(2)-}+i\varepsilon \right) },
  \label{pm}
\end{equation}
and just for the purpose of comparison we quote the result for the
{\footnotesize$(\perp \perp ,\perp \perp)$}-component,
\begin{equation}
S^{(2)\perp \perp ,\perp \perp }(P^{-})=\frac{\theta (k_{1}^{+})\theta
(P^{+}-k_{1}^{+})}{2k_{1}^{+}\left( P^{+}-k_{1}^{+}\right) }\frac{i\left(
-g^{\perp \perp }\right) \left( -g^{\perp \perp }\right) }{\left(
P^{-}-K_{0}^{(2)-}+i\varepsilon \right) },  \label{pp}
\end{equation}
where 
\begin{equation}
K_{0}^{(2)-}\equiv k_{1\rm on}^{-}+k_{2\rm on}^{-}=\frac{k_{1}^{\perp 2}}{k_{1}^{+}}+\frac{(P^\perp -k_1^\perp)^{2}}{ P^{+}-k_{1}^{+}},
\label{k02}
\end{equation}
with 
\[
k_{i\,\rm{on}}^-=\frac{k_i^{\perp 2}}{k_i^{+}}\qquad i=1,2.
\].

$K_{0}^{(2)-}$ is the light-front Hamiltonian of the free two-particle
system. For $x^{+}\ <\ 0$, $S^{(2)}(x^{+})\ =\ 0$ due to our choice of
$ P^{+}>0$. Observe that $S^{(2)}(P^{-})$ is written in Eq.(\ref{pm})
and Eq.( \ref{pp}) in operator form with respect to $k_1^{+}$ and
$\vec{k}_1^{\perp }$.  

We have written down explicitly the result for
$S^{(2)\perp\perp,\perp\perp}$ in order to point out that this
component is free from the zero mode, but as in the one-boson case
before, again we have problems of a divergent factor for $k_{1}^{+}=0$
in (\ref{pm}), which is a manifestation of the zero mode problem
lingering on even after the ML prescription has been used. Other
non-vanishing components, namely, $S^{(2)--,--}$, $S^{(2)--,-\perp}$,
$S^{(2)--,\perp\perp}$ and $S^{(2)-\perp,\perp\perp}$ can be evaluated
in a similar fashion and all of these also contain zero mode
problems.

\section{Zero modes through the singularity-softening prescription}

Zero modes in the light front milieu is a very subtle problem which
for years have been challenging us with the best of our efforts to
understand it, to make it manageable and to make some physical sense
out of it. We have already learned that a prescription to handle those
singularities cannot be solely mathematically well-defined --- that is
not enough --- we now know that the prescription must be causal, that
is, it needs to ascertain that its implementation does not violate
causality \cite{cp}. ML prescription has been heralded as the causal
prescription to handle the light-front singularites and many a
calculation do confirm that it has solved many difficulties concerning
one- and two-loop quantum corrections to Feynman diagrams. However, as
seen in the previous sections, ML prescription does not remove the
pathological zero modes in the one- and two-vector gauge boson
propagation at the quantum level. We therefore come to the place most
important in this work: The introduction of a novel prescription that
is causal and can handle the light-front singularities which does not
leave remnant zero modes is presented and applied to the one and two
propagating vector bosons in the light-front gauge.

The index {\rm SS} stands for this singularity-softening prescription for the
treatment of the $(k^{+})^{-1}$ poles (cf.\cite{covar}), namely, 
\begin{eqnarray}\label{ss}
\left[ \frac{1}{k^{+}}\right] _{{\rm SS}} &=&\lim_{\varepsilon \rightarrow 0}
\left[ \frac{k^{2}}{k^{+}\left( k^{2}+i\varepsilon \right) }\right] _{{\rm SS
}}  \nonumber \\
&=&\lim_{\varepsilon \rightarrow 0}\left[ \frac{k^{-}-k_{{\rm on}}^{-}}{
k^{+}\left( k^{-}-k_{{\rm on}}^{-}+\frac{i\varepsilon }{k^{+}}\right) }
\right] _{{\rm SS}} 
\end{eqnarray}

\section{The one gauge boson case}

The component $S^{\perp -}$ is: 
\begin{eqnarray}
S^{(1)\perp -}(p^{-}) &=&i\int dk_{1}^{-}\frac{k_{1}^{\perp }\;n^{-}\delta
\left( p^{-}-k_{1}^{-}\right) }{k_{1}^{+}\left( k_{1}^{-}-\frac{k_{1}^{\perp
2}}{k_{1}^{+}}+\frac{i\varepsilon }{k_{1}^{+}}\right) }\left[ \frac{1}{
k_{1}^{+}}\right] _{{\rm SS}}  \nonumber \\
&=&i\int dk_{1}^{-}\frac{k_{1}^{\perp }\;n^{-}\delta \left(
p^{-}-k_{1}^{-}\right) }{k_{1}^{+}\left( k_{1}^{-}-\frac{k_{1}^{\perp 2}}{
k_{1}^{+}}+\frac{i\varepsilon }{k_{1}^{+}}\right) }  \nonumber \\
&&\times \left[ \frac{k_{1}^{-}-k_{1{\rm on}}^{-}}{k_{1}^{+}\left(
k_{1}^{-}-k_{1{\rm on}}^{-}+\frac{i\varepsilon }{k_{1}^{+}}\right) }\right]
_{{\rm SS}}
\end{eqnarray}

The result is: 
\begin{equation}
S^{(1)\perp -}(p^{-})=\frac{\theta (p^{+})\;p^{\perp }n^{-}}{p^{+}}\left[ 
\frac{p^{-}-p_{{\rm on}}^{-}}{p^{+}\left( p^{-}-p_{{\rm on}}^{-}+\frac{
i\varepsilon }{p^{+}}\right) }\right] _{{\rm SS}}\frac{i}{\left(
p^{-}-K_{0}^{(1)-}+i\varepsilon \right) },
\end{equation}
where we have introduced the definition 
\begin{equation}
K_{0}^{(1)-}=p_{{\rm on}}^{-}=\frac{p^{\perp 2}}{p^{+}},
\end{equation}

However, since the Dirac delta funtion $\delta \left(
p^{-}-k_{1}^{-}\right) $ forces us onto the mass-shell, the numerator
is identically zero, that is,
\[
p^{-}-p^-_{\rm on}=0 
\]
and this is true for massless as well as massive gauge bosons.

Thus, finally
\[
S^{(1)\perp -}(p^{-})=0. 
\]

For the component $S^{--}$ we have: 
\begin{eqnarray}
S^{(1)--}(p^{-}) &=&i\int dk_{1}^{-}\frac{k_{1}^{\perp 2}\,n^{-}n^{-}\delta
\left( p^{-}-k_{1}^{-}\right) }{k_{1}^{+}\left( k_{1}^{-}-\frac{k_{1}^{\perp
2}}{k_{1}^{+}}+\frac{i\varepsilon }{k_{1}^{+}}\right) }\left[ \frac{1}{
\left( k_{1}^{+}\right) ^{2}}\right] _{{\rm SS}}  \nonumber \\
&=&i\int dk_{1}^{-}\frac{k_{1}^{\perp 2}n^{-}n^{-}\delta \left(
p^{-}-k_{1}^{-}\right) }{k_{1}^{+}\left( k_{1}^{-}-\frac{k_{1}^{\perp 2}}{
k_{1}^{+}}+\frac{i\varepsilon }{k_{1}^{+}}\right) }  \nonumber \\
&&\times \left[ \frac{k_{1}^{-}-k_{1{\rm on}}^{-}}{k_{1}^{+}\left(
k_{1}^{-}-k_{1{\rm on}}^{-}+\frac{i\varepsilon }{k_{1}^{+}}\right) }\right]
_{{\rm SS}}^{2}
\end{eqnarray}

which results in: 
\begin{equation}
S^{(1)--}(p^{-})=\frac{\theta (p^{+})\;p^{\perp 2}n^{-}n^{-}}{p^{+}}
\left[ \frac{p^{-}-p_{{\rm on}}^{-}}{p^{+}\left( p^{-}-p_{{\rm on}}^{-}+
\frac{i\varepsilon }{p^{+}}\right) }\right] _{{\rm SS}}^{2}\frac{i}{
\left( p^{-}-K_{0}^{(1)-}+i\varepsilon \right) }.
\end{equation}

For the same reason as before stated,  
\[
S^{(1)--}(p^{-})=0 
\]

In the case of $S^{(1)\perp \perp }(p^{-})$ component we have 
\[
S^{(1)\perp \perp }(P^{-})=\frac{\theta (p^{+})}{p^{+}}\frac{i\left(
-g^{\perp \perp }\right) }{\left( p^{-}-K_{0}^{(1)-}+i\varepsilon \right) } 
\]

Clearly, this case does not present us with the $p^{+}=0$ difficulty,
and the only non-vanishing result is just $S^{(1)\perp \perp }$. Only
the physical degrees of freedom (transverse ones) do propagate and
without zero mode hindrances anywhere!

\section{The two gauge bosons case}

Let us go component by component in the possible non-vanishing
contributuions as given in (\ref{2boson}). For $S^{(2)\perp -,\perp
-}$ we have
\begin{eqnarray}
S^{(2)\perp -,\perp -}(P^{-}) &=&-\frac{1}{\left( 2\pi \right) }\int \frac{
dk_{1}^{-}}{k_{1}^{+}\left( P^{+}-k_{1}^{+}\right) } \\
&\times &\left[ \frac{k_{1}^{-}-k_{1{\rm on}}^{-}}{k_{1}^{+}\left(
k_{1}^{-}-k_{1{\rm on}}^{-}+\frac{i\varepsilon }{k_{1}^{+}}\right) }\right]
_{{\rm SS}}\frac{n^{-}k_{1}^{\perp }}{\left( k_{1}^{-}-\frac{k_{1}^{\perp 2}
}{k_{1}^{+}}+\frac{i\varepsilon }{k_{1}^{+}}\right) }  \nonumber \\
&\times &\left[ \frac{P^{-}-k_{1}^{-}-k_{2{\rm on}}^{-}}{\left(
P^{+}-k_{1}^{+}\right) \left( P^{-}-k_{1}^{-}-k_{2{\rm on}}^{-}+\frac{
i\varepsilon }{P^{+}-k_{1}^{+}}\right) }\right] _{{\rm SS}}  \nonumber \\
&\times &\frac{n^{-}k_{2}^{\perp }}{\left( P^{-}-k_{1}^{-}-\frac{\left(
P^{\perp }-k_{1}^{\perp}\right)^{2}}{P^+-k_{1}^{+}}+\frac{i\varepsilon }{P^{+}-k_{1}^{+}}\right) }\ .  \nonumber
\end{eqnarray}

Evaluating the residue at the pole
\begin{equation}\label{pole}
k_{1}^{-}=k_{1{\rm on}}^{-}-\frac{i\varepsilon }{k_{1}^{+}}\text{ ,} 
\end{equation}
we have
\[
S^{(2)\perp -,\perp -}(P^{-})=0
\]

For the $S^{(2)--,--}$-component we have
\begin{eqnarray}
S^{(2)--,--}(P^{-}) &=&-\frac{1}{\left( 2\pi \right) }\int \frac{dk_{1}^{-}}{
k_{1}^{+}\left( P^{+}-k_{1}^{+}\right) } \\
&\times &\left[ \frac{k_{1}^{-}-k_{1{\rm on}}^{-}}{k_{1}^{+}\left(
k_{1}^{-}-k_{1{\rm on}}^{-}+\frac{i\varepsilon }{k_{1}^{+}}\right) }\right]
_{{\rm SS}}^{2}\frac{k_{1}^{\perp 2}\,n^{-}n^{-}}{\left( k_{1}^{-}-\frac{
k_{1}^{\perp 2}}{k_{1}^{+}}+\frac{i\varepsilon }{k_{1}^{+}}\right) } 
\nonumber \\
&\times &\left[ \frac{P^{-}-k_{1}^{-}-k_{2{\rm on}}^{-}}{\left(
P^{+}-k_{1}^{+}\right) \left( P^{-}-k_{1}^{-}-k_{2{\rm on}}^{-}+\frac{
i\varepsilon }{P^{+}-k_{1}^{+}}\right) }\right] _{{\rm SS}}^{2}  \nonumber \\
&\times &\frac{k_{2}^{\perp 2}\,n^{-}n^{-}}{\left( P^{-}-k_{1}^{-}-\frac{
\left( P^{\perp }-k_{1}^{\perp}\right)^{2}}{P^+-k_{1}^{+}}+\frac{i\varepsilon }{P^{+}-k_{1}^{+}}\right) }\ .  \nonumber
\end{eqnarray}
yielding also
\[
S^{(2)--,--}(P^{-})=0
\]

For the component $S^{(2)\perp -,--}$ we have
\begin{eqnarray}
S^{(2)\perp -,--}(P^{-}) &=&-\frac{1}{\left( 2\pi \right) }\int \frac{
dk_{1}^{-}}{k_{1}^{+}\left( P^{+}-k_{1}^{+}\right) } \\
&\times &\left[ \frac{k_{1}^{-}-k_{1{\rm on}}^{-}}{k_{1}^{+}\left(
k_{1}^{-}-k_{1{\rm on}}^{-}+\frac{i\varepsilon }{k_{1}^{+}}\right) }\right]
_{{\rm SS}}\frac{k_{1}^{\perp }\,n^{-}}{\left( k_{1}^{-}-\frac{k_{1}^{\perp 2}
}{k_{1}^{+}}+\frac{i\varepsilon }{k_{1}^{+}}\right) }  \nonumber \\
&\times &\left[ \frac{P^{-}-k_{1}^{-}-k_{2{\rm on}}^{-}}{\left(
P^{+}-k_{1}^{+}\right) \left( P^{-}-k_{1}^{-}-k_{2{\rm on}}^{-}+\frac{
i\varepsilon }{P^{+}-k_{1}^{+}}\right) }\right] _{{\rm SS}}^{2}  \nonumber \\
&\times &\frac{k_{2}^{\perp 2}\,n^{-}n^{-}}{\left( P^{-}-k_{1}^{-}-\frac{
\left( P^{\perp }-k_{1}^{\perp}\right) ^{2}}{P^+-k_{1}^{+}}+\frac{i\varepsilon }{P^{+}-k_{1}^{+}}\right) }\ .  \nonumber
\end{eqnarray}
which results in
\[
S^{(2)\perp -,--}(P^{-})=0
\]

For the component $S^{(2)\perp -,\perp \perp }$ we have
\begin{eqnarray}
S^{(2)\perp -,\perp \perp }(P^{-}) &=&-\frac{1}{\left( 2\pi \right) }\int 
\frac{dk_{1}^{-}}{k_{1}^{+}\left( P^{+}-k_{1}^{+}\right) } \\
&\times &\left[ \frac{k_{1}^{-}-k_{1{\rm on}}^{-}}{k_{1}^{+}\left(
k_{1}^{-}-k_{1{\rm on}}^{-}+\frac{i\varepsilon }{k_{1}^{+}}\right) }\right]
_{{\rm SS}}\frac{k_{1}^{\perp }\,n^{-}}{\left( k_{1}^{-}-\frac{k_{1}^{\perp 2}
}{k_{1}^{+}}+\frac{i\varepsilon }{k_{1}^{+}}\right) }  \nonumber \\
&\times &\frac{\left( -g^{\perp \perp }\right) }{\left( P^{-}-k_{1}^{-}-
\frac{\left( P^{\perp }-k_{1}^{\perp}\right) ^{2}}{P^+-k_{1}^{+}}+\frac{i\varepsilon }{P^{+}-k_{1}^{+}}\right) }\ .  \nonumber
\end{eqnarray}
leading us to he same vanishing result
\[
S^{(2)\perp-,\perp\perp}(P^-)=0
\]

The component $S^{(2)--,\perp \perp }$ is given by
\begin{eqnarray}
S^{(2)--,\perp \perp }(P^{-}) &=&-\frac{1}{\left( 2\pi \right) }\int \frac{
dk_{1}^{-}}{k_{1}^{+}\left( P^{+}-k_{1}^{+}\right) } \\
&\times &\left[ \frac{k_{1}^{-}-k_{1{\rm on}}^{-}}{k_{1}^{+}\left(
k_{1}^{-}-k_{1{\rm on}}^{-}+\frac{i\varepsilon }{k_{1}^{+}}\right) }\right]
_{{\rm SS}}^{2}\frac{k_{1}^{\perp 2}\,n^{-}n^{-}}{\left( k_{1}^{-}-\frac{
k_{1}^{\perp 2}}{k_{1}^{+}}+\frac{i\varepsilon }{k_{1}^{+}}\right) } 
\nonumber \\
&\times &\frac{\left( -g^{\perp \perp }\right) }{\left( P^{-}-k_{1}^{-}-
\frac{\left( P^{\perp }-k_{1}^{\perp}\right) ^{2}}{P^+-k_{1}^{+}}+\frac{i\varepsilon }{P^{+}-k_{1}^{+}}\right) }\ .  \nonumber
\end{eqnarray}
which also yields
\[
S^{(2)--,\perp\perp}(P^-)=0
\]

Finally, for the component $S^{(2)\perp \perp ,\perp \perp }$ we have
\begin{eqnarray}
S^{(2)\perp \perp ,\perp \perp }(P^{-}) &=&-\frac{1}{\left( 2\pi \right) }
\int \frac{dk_{1}^{-}}{k_{1}^{+}\left( P^{+}-k_{1}^{+}\right) }\frac{\left(
-g^{\perp \perp }\right) }{\left( k_{1}^{-}-\frac{k_{1}^{\perp 2}}{k_{1}^{+}}
+\frac{i\varepsilon }{k_{1}^{+}}\right) }  \nonumber \\
&\times &\frac{\left( -g^{\perp \perp }\right) }{\left( P^{-}-k_{1}^{-}-
\frac{\left( P^{\perp }-k_{1}^{\perp}\right) ^{2}}{P^+-k_{1}^{+}}+\frac{i\varepsilon }{P^{+}-k_{1}^{+}}\right) }\ .  \nonumber
\end{eqnarray}
which yields the non-vanishing contribution
\[
S^{(2)\perp \perp ,\perp \perp }(P^{-})=\frac{\theta (k_{1}^{+})\theta
(P^{+}-k_{1}^{+})}{k_{1}^{+}\left( P^{+}-k_{1}^{+}\right) }\frac{i\left(
-g^{\perp \perp }\right) \left( -g^{\perp \perp }\right) }{\left(
P^{-}-K_{0}^{(2)-}+i\varepsilon \right) },
\]
which is the same as that obtained through the ML-prescription, (\ref{pp}).

Therefore, as long as we treat the troublesome zero modes $k^+=0$ via
the singularity-softening prescription (\ref{ss}) the only
non-vanishing component of the two gauge boson propagator is the
{\footnotesize$(\perp\!\perp,\perp\!\perp)$}-component, so that there
is no zero mode problem left and the only propagating modes are the
physical, transverse ones!

\section{Conclusion}

Projecting the Feynman covariant space propagator into the light-front
coordinates and using the Mandelstam-Leibbrandt prescription to treat
the $k^+=0$ singularities we got propagation of one and two bodies in
the light-front. However, the result is beset with zero-mode
pathology, for the one gauge boson case and also for the two boson
case in non-vanishing components such as {\footnotesize $(\perp\! -,
\perp\! -)$}; the only exception being the {\footnotesize $(\perp
\perp, \perp \perp)$}-component where there is no singularity of this
type. We observe that even with the use of the Leibbrandt-Mandelstam
prescription, it was not possible to remove the built-in singularity
in $k^+=0$ --- the outstanding zero mode problem of light-front.

Now, with the introduction of a causality preserving and
singularity-softening prescription, we were able to remove the zero
mode problem and let only the physical degrees of freedom propagate,
namely, the transverse ones, not only in one gauge boson propagation
but also in the propagation of two gauge bosons.
\vspace{.5cm}

{\bf Acknowledgments:} A.T.S. wishes to thank the kind hospitality of
Physics Department, North Carolina State University, Raleigh, NC and
acknowledges research grant in the earlier part of this work from CNPq
(Bras\'{i}lia, DF), superseded by a grant from CAPES (Bras\'{\i}lia,
DF). J.H.O.S. thanks for the hospitality of Instituto de F\'{\i}sica
Te\'orica-UNESP, which provided facilities for the completion of this
work.

\end{document}